\documentclass[12pt,cite,article]{iopart}

\usepackage{graphicx}
\usepackage[hang,nooneline]{subfigure}
\usepackage{cite}

\newcommand{\degr}{$^{\rm{o}}$}

\begin{document}

\title{Graphene on Ru(0001): A corrugated and chiral structure}

\author{D.~Martoccia$^1$, M.~Bj\"orck$^{1}$\footnote{Present address: MAX-lab, P.O. Box 118, SE-22100 Lund, Sweden.}, C.M.~Schlep\"utz$^{1}$\footnote{Present address: Physics Department, University of Michigan, Ann Arbor, Michigan 48109-1120, USA}, T.~Brugger$^2$, S.A.~Pauli$^1$, B.D.~Patterson$^1$, T.~Greber$^2$ and P.R.~Willmott$^1$}

\address{$^1$Swiss Light Source, Paul Scherrer Institut, CH-5232 Villigen, Switzerland.}
\address{$^2$Institute of Physics, University of Z\"urich, Winterthurerstrasse 190, CH-8057 Z\"urich, Switzerland.}

\ead{philip.willmott@psi.ch}

%
%
%
%
%
%


\date{\today}

\begin{abstract}
We present a structural analysis of the graphene/Ru(0001) system obtained by surface x-ray diffraction. The data were fit using Fourier-series expanded displacement fields from an ideal bulk structure, plus the application of symmetry constraints. The shape of the observed superstructure rods proves a reconstruction of the substrate, induced by strong bonding of graphene to ruthenium. Both the graphene layer and the underlying substrate are corrugated, with peak-to-peak heights of $(0.82\pm0.15)$~\AA\ and $(0.19\pm0.02)$~\AA\ for the graphene and topmost Ru-atomic layer, respectively. The Ru-corrugation decays slowly over several monolayers into the bulk. The system also exhibits chirality, whereby in-plane rotations of up to $2.0$\degr in those regions of the superstructure where the graphene is weakly bound are driven by elastic energy minimization.
\end{abstract}

\maketitle

\section{Introduction}
Nanostructured materials have attracted increasing interest in recent years, due to their potential in practical electronic applications. One of these, graphene, has been theoretically investigated since the 1940s \cite{Wallace1947}. The discovery in 2004 that freestanding graphene may be prepared \cite{Novoselov2004} led to an explosion of interest in this material due to its unique electronic properties and possible practical utilization \cite{Geim2007}. Graphene is a single sheet of carbon atoms arranged in a honeycomb structure, which was believed to be thermodynamically unstable under ambient conditions, due to the Mermin-Wagner theorem \cite{Mermin1966}. Nowadays the stability of graphene is explained by postulating small out-of-plane corrugations, leading to lower thermal vibrations \cite{Meyer2007, Fasolino2007}. 
On crystalline substrates the formation of superstructures and the concomitant corrugation of graphene provides template functionality \cite{Brugger2009}. The influence of the substrate and the formation of the superstructure is believed to change the electronic bandstructure and the electronic properties \cite{Wehling2008,Park2008}, due to bond formation and charge-transfer phenomena \cite{Sutter2008, Giovannetti2008, Isacsson2008, Brugger2009}. The characterization of the graphene--metal interface structure is of crucial importance, because measurements of the electronic transport properties require making metallic contacts \cite{Ishigami2007}. Surface x-ray diffraction (SXRD) is a powerful investigative tool for this system, since the diffraction intensity is perfectly described in a single scattering picture and is unaffected by density-of-states effects or electrostatic forces.

Graphene grown on transition metals forms single-domain superstructures with high degrees of structural perfection \cite{Marchini2007b, Pan2007c, deParga2008a, Martoccia2008a, Pan2008}. Early reports on graphene on Ru(0001) proposed a superstructure in which $(12 \times 12)$ unit cells of graphene sit on $(11 \times 11)$ unit cells of ruthenium, (``12-on-11'') \cite{Marchini2007b, Pan2007c}, while other studies proposed an 11-on-10 superstructure \cite{deParga2008a}. However, recent SXRD  results showed unambiguously that the reconstruction is in fact a surprisingly large 25-on-23 superstructure \cite{Martoccia2008a}. A comparative study between density functional theory (DFT) calculations and scanning tunneling microscopy (STM) experiments showed the structure to be composed of regions of alternating weak and strong chemical interactions of the graphene with the Ru-substrate \cite{Wang2008b, Jiang2009}.

Here, we detail the atomic structure of the graphene/Ru(0001) system, determined with sub-Angstrom resolution from SXRD data.  In addition to quantifying the corrugation, we also show that the best model exhibits the formation of chiral domains, resulting in a lower symmetry ($p3$) compared to graphite ($p3m1$). This unexpected property may have an important impact on e.g., the use of this system as a template for molecular chiral recognition. We argue that this breaking of the symmetry is driven by energy minimization based on elastic energy considerations.

\section{Experimental}
Sample preparation and the SXRD measurement setup at the Surface Diffraction station of the Materials Science beamline, Swiss Light Source, have already been detailed in \cite{Martoccia2008a}. It was demonstrated from simple simulations of the $25/23$ superstructure rod (SSR) that the substrate must also be corrugated, since oscillations with the appropriate periodicity of approximately $1.0$~out-of-plane substrate reciprocal lattice units (r.l.u., $2\pi/c$) on the SSRs only start to appear if one includes a corrugation of the substrate. Here, we present further SXRD data from the same sample, which in addition to the SSRs now also includes in-plane data.

Because of the very large number of atoms involved in the superstructure, it is impossible to fit each atomic position individually. Instead, we have parametrized the structural model using a small set of physically reasonable parameters. The in-plane and out-of-plane deviations of the atomic positions from an ideal flat structure of the graphene and of the uppermost layers of the Ru-substrate are described by a 2D Fourier-series expansion. We truncate this series after the fourth Fourier component, since higher orders could not be resolved in the diffraction data. The displacement field of the system is allowed to adopt the lower $p3$ symmetry, since this is the lowest symmetry still compatible with the \emph{apparent} measured sixfold diffraction symmetry, which only arises because of the superposition of the two possible terminations of the hexagonal close-packed (hcp) substrate \cite{delaFiguera2006}. Because the $p3$ symmetry allows chiral structures, we have to sum over the signals from domains of each enantiomer, and assume a $50$~$\%$ distribution.

Details of the implementation of the Fourier expansion and of symmetry constraints are given in the Appendix. Here, we only discuss those aspects needed to understand the results. First, it is important to note that because the 25-on-23 structure contains $2\times 2$ corrugation periods, only the even Fourier components, that is the second and fourth, must be considered. This is also demonstrated by the absence of signal at the $22/23$, $24/23$,~\ldots \, superstructure rods. For each atom within the supercell, the in-plane and out-of-plane deviations $\Delta x$, $\Delta y$, and $\Delta z$ are described by the two Fourier components. In total, both the graphene and ruthenium require nine fitting parameters each in order to describe their corrugations.

In addition to the $18$ corrugation parameters we introduce a factor, $\lambda$, which describes an exponential decay of the substrate corrugation amplitude with substrate depth $z$
\begin{eqnarray}
      A(z) & = & A_{0} \cdot \exp(-z/\lambda).
\end{eqnarray}
This decay applies to all the three amplitudes used for the description of the substrate displacement function. We fix the minimum distance from the substrate to the graphene layer, $d_{\mathrm{C-Ru}}$, to $2.0$~\AA\ \cite{Wang2008b, Sutter2009b} since our model is relatively insensitive to this parameter within physically sensible limits  ($\pm0.1$~\AA) . The parameter $d_{\mathrm{Ru_{1}-Ru_{2}}}$ is the distance between the first and second Ru-atomic layer. Lastly, a global scaling factor $S$ is required, resulting in a total of $21$~free fitting parameters.

\begin{figure}
  \includegraphics[width=1.0\columnwidth]{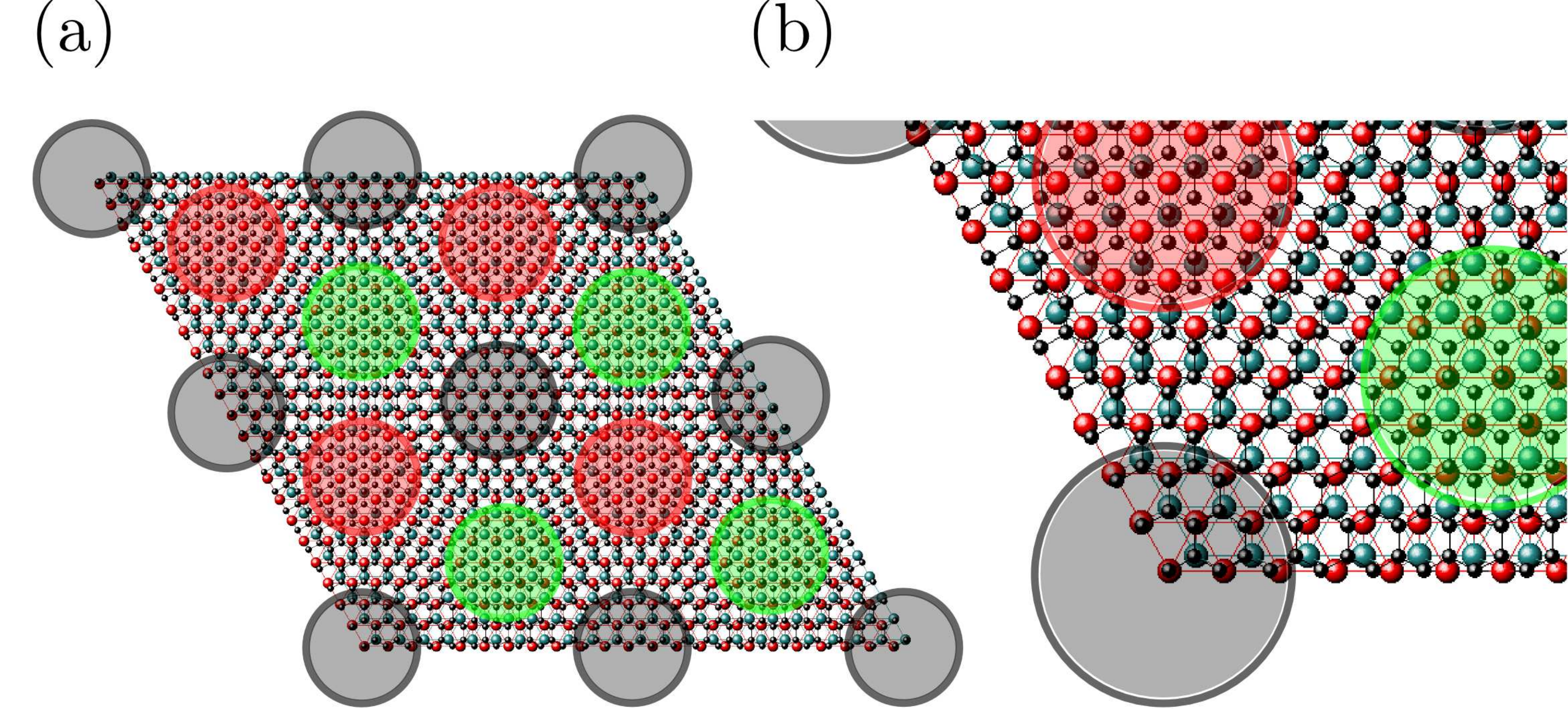}
  \caption{(color) (a) Graphene (black) on top of ruthenium (red, first layer) and green (second layer). Three regions shaded gray, green, and red, are highlighted, and explained in the text. (b) A zoom into the lower left corner of the $25$-on-$23$ supercell.} 
  \label{figure:flat_structure}
\end{figure}

We begin by defining regions of the supercell, where we consider a flat graphene layer lying commensurably 25-on-23 on top of a flat Ru-substrate (see Fig.~\ref{figure:flat_structure}). The gray-shaded region in Fig.~\ref{figure:flat_structure}(b) indicates where the first of the two C-atoms within a `normal' graphene unit cell sits on top of a Ru-atom of the topmost substrate atomic layer (red atoms), and the second atom sits on top of a Ru-atom from the second substrate atomic layer (green atoms, the hcp position). Henceforth, we refer to this as the (top,hcp) region. Using the same arguments, the red area is the (hcp,fcc) region, and the green is the (fcc,top) \cite{Grad2003}. 

Fitting \cite{Errorfunction2009} was performed using GenX \cite{Bjorck2007}, an optimization program using the differential evolution algorithm, which helps avoid getting trapped in local minima \cite{Wormington1999}. The errors on the fitted parameters are estimated by an increase in the goodness-of-fit of $5$~\%.

\begin{figure}
  \includegraphics[width=1.0\columnwidth]{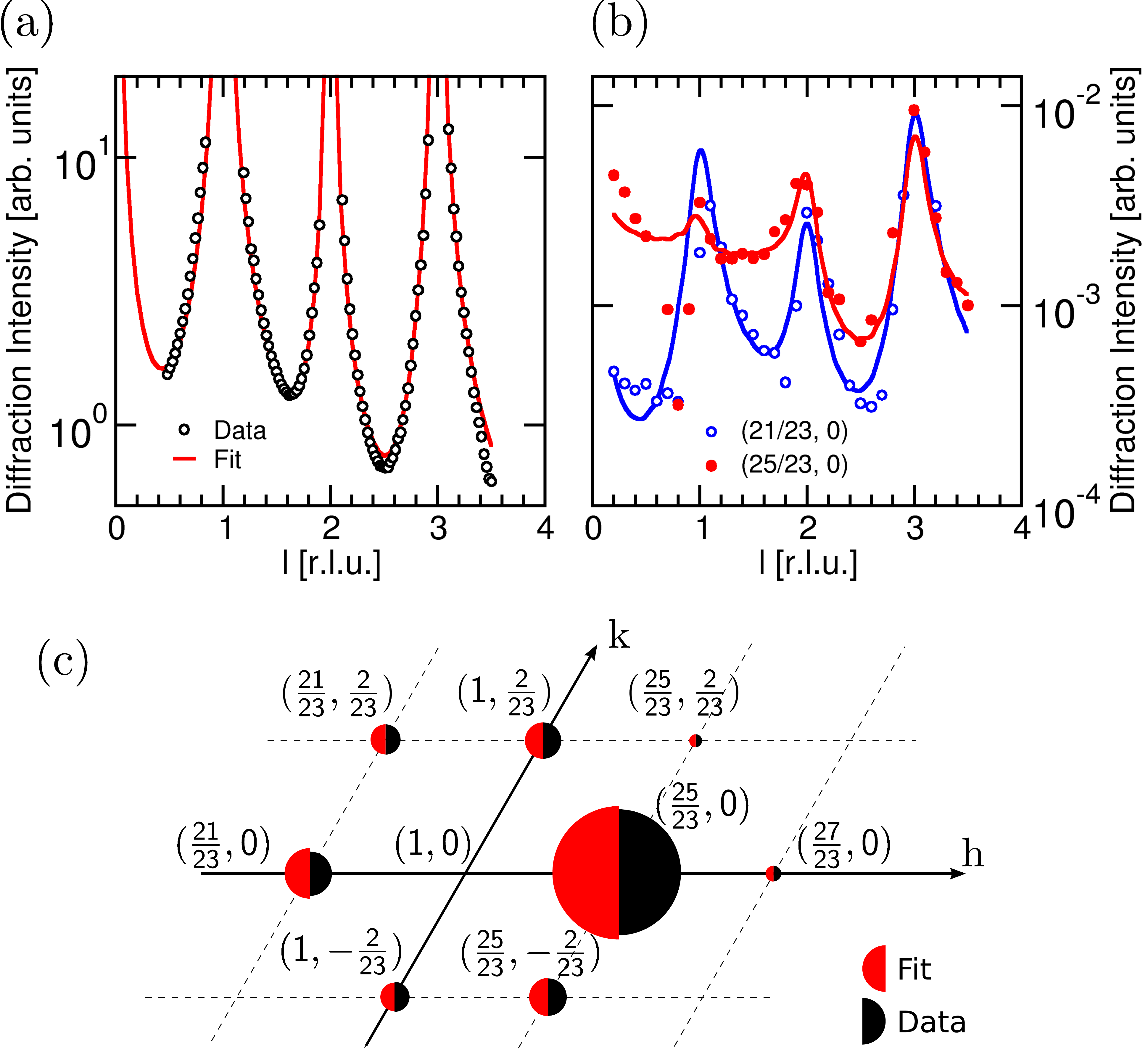}
  \caption{(color online) (a) The (1,0)-CTR. Only the scaling factor and $d_{\mathrm{Ru_{1}-Ru_{2}}}$ were used to fit the data. (b) Fit of the two superstructure rods. (c) In-plane map of the superstructure reflections around the (1,0)-CTR position at $l=0.4$~r.l.u. The areas of the circles are proportional to the scattering intensities.}
  \label{figure:data}
\end{figure}

We fit $d_{\mathrm{Ru_{1}-Ru_{2}}}$ to the CTR-data alone [Fig.~\ref{figure:data}(a)] as this is sensitive to small differences in the interplanar spacing of the topmost two Ru-atomic layers but is largely insensitive to the form of the weakly scattering superstructure. The best fit had an R-factor of $5.2$~\%, for  $d_{\mathrm{Ru_{1}-Ru_{2}}} = (2.080\pm0.003)$~\AA, which should be compared to a bulk value of $2.141$~\AA. This equates to a contraction of $2.8$~\%, in agreement with \cite{Michalk1983, Feibelman1994, Baddorf2002}.

\section{Results and discussion}
\begin{figure}
  \includegraphics[width=1.0\columnwidth]{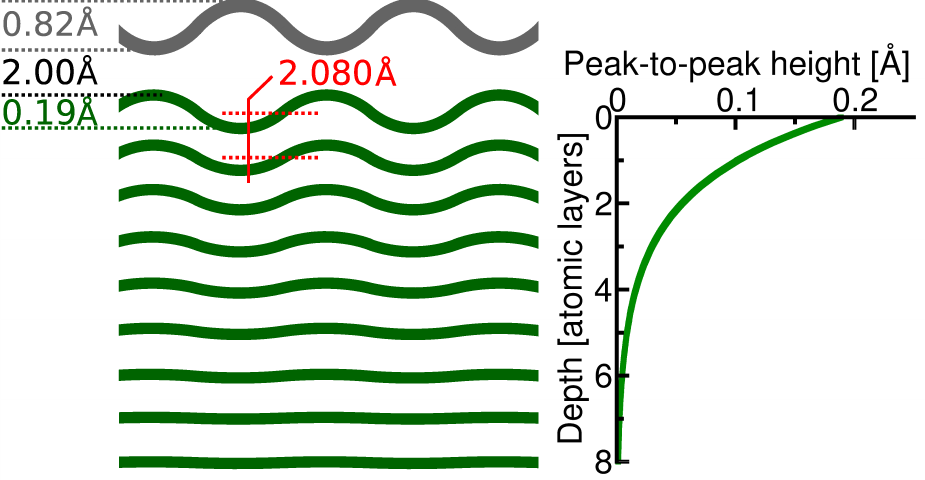}
  \caption{(color online) Schematic view of the corrugation and interplanar distances of the graphene and substrate.}
  \label{figure:antiphase}
\end{figure}

The starting model for the search of all the other parameters was a strained 25-on-23 flat graphene layer lying commensurably on a flat ruthenium bulk structure. The best fit for the SSR- and in-plane data has an R-factor of $13.4$~\% [Fig.~\ref{figure:data}(b) and (c)]. The peak-to-peak corrugation height of the graphene is $(0.82\pm0.15)$~\AA, in agreement with \cite{Marchini2007b, deParga2008a, Sutter2009}, whereas that of the uppermost Ru-atomic layer is $(0.19\pm0.02)$~\AA\ and is out of phase with respect to the graphene corrugation (Fig.~\ref{figure:antiphase}). The exponential decay length of $\lambda=(7.0\pm0.4)$~\AA\ means there is still approximately a tenth of the distortion of the first Ru-atomic layer at a depth of four Ru-atomic layers.  This strongly supports the idea of a chemisorbed graphene layer with significant interaction with the substrate \cite{Preobrajenski2008, Sutter2009b, McCarty2009, Sun2009, Brugger2009}. 

\begin{figure}
\begin{center}
  \includegraphics[width=0.7\columnwidth]{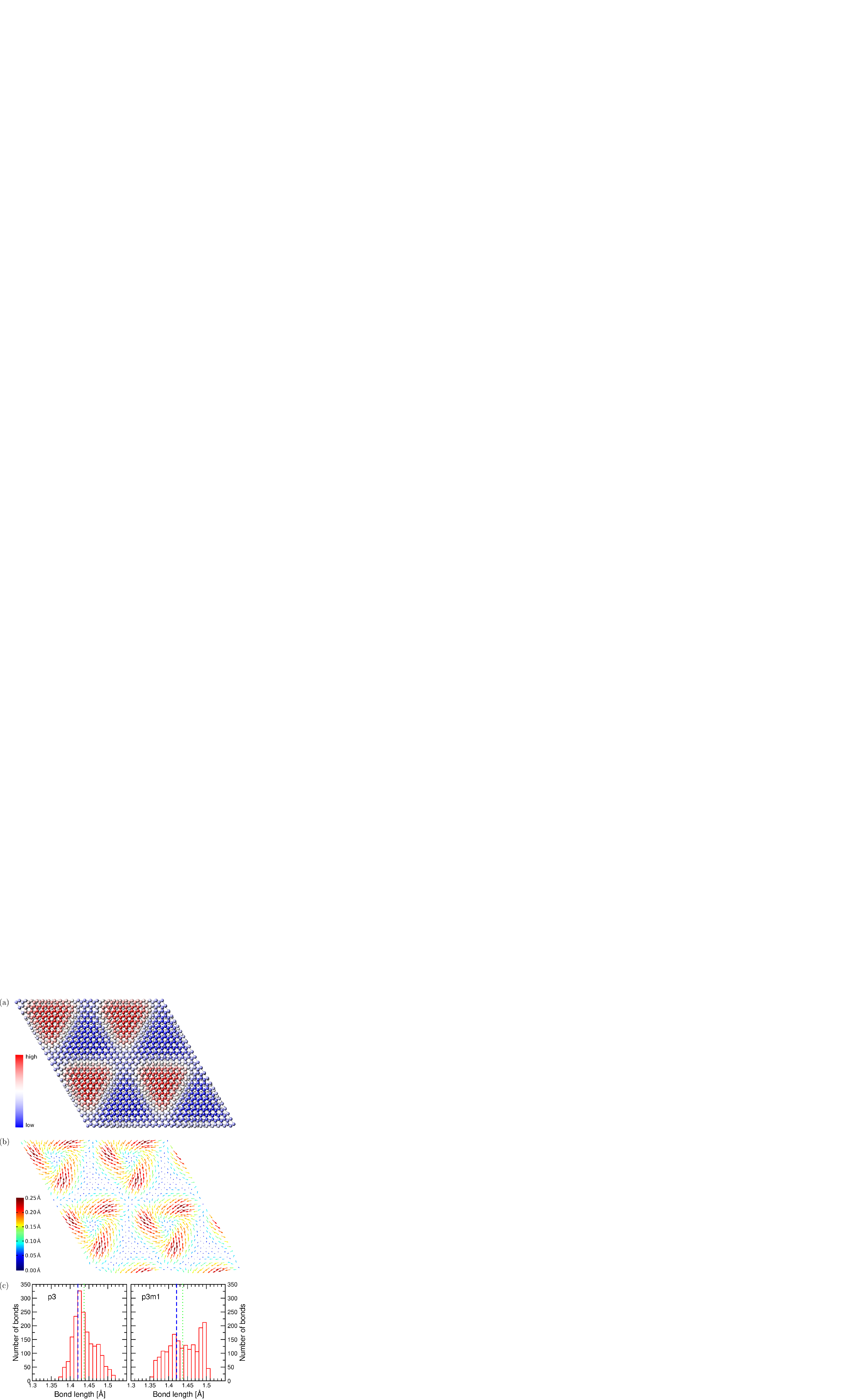}
  \caption{(color) (a) Top view of the counterclockwise twisting enantiomer resulting from the fitting procedure: The graphene shows the lowest lying atoms to be in (top,hcp)-region, whereas the hill maxima appear in the (hcp,fcc)-regions. Clear triangular-shaped hills are observed. (b) The in-plane displacements of the same enantiomer, magnified by a factor of $10$, from the ideal bulk positions. The distortions are largest on the flanks of the hills. (c) Histogram of the bond lengths in the graphene layer. The model with $p3$~symmetry allows the carbon hexagons to twist, and most of the bonds are stretched by less than $\pm 0.04$~\AA\ compared to the bulk bond length of graphite ($1.421$~\AA) (blue, dashed line) or a flat $25/23$ superstructure bond length ($a_{\mathrm{Ru}}/\sqrt{3} \times 23/25 = 1.4373$~\AA) (green, dotted line). Enforcing the higher $p3m1$-symmetry causes larger distortions in the bond lengths.}
\label{figure:inplanemovements3rows}
\end{center}
\end{figure}

Details of the final structure are summarized in Fig.~\ref{figure:inplanemovements3rows}. Figure~\ref{figure:inplanemovements3rows}(a) shows a clear corrugation of the graphene with the hills lying in the weakly bound (hcp,fcc)-region. The hills have a triangular shape, in remarkable agreement with earlier STM data \cite{Marchini2007b, Pan2007c}. Although in-plane \emph{movements} of up to $(0.25\pm0.03)$~\AA\ of the graphene are observed [Fig.~\ref{figure:inplanemovements3rows}(b)], the bond \emph{lengths} are distorted by less than $0.1$~\AA. This requires a twisting motion and indeed the in-plane movements exhibit a chiral signature, in which the largest movements occur at the steepest flanks of the hills, as one might expect, based on simple elastic strain considerations. Note that this feature emerged naturally from the fitting and was not implemented a~priori into the model. The biggest rotation angle of the hexagons is $2.0$\degr\, found on the flanks as well as on top of the hills. 

The elastic energy was calculated to test the physical validity of the presented parametrization approach and the resulting model. It takes into account the in-plane and out-of-plane displacements of surface atoms from their `ideal' positions due to the $25/23$ surface reconstruction. From our model, we calculate an elastic energy \cite{Keating1966, Pedersen1989, Bunk1999} due to strain of $9.3$~eV per supercell, assuming zero strain for a flat 25-on-23 graphene layer \cite{LatticeConstantFootnote2010}. Fitting the data to the higher $p3m1$-symmetry results in an increase in the elastic energy by $83$~\%, while the R-factor of $14.7$~\% is significantly higher than that for the $p3$-symmetry. Even if we were to assume zero strain for a flat graphene layer having the bulk graphite in-plane lattice constant [Fig.~\ref{figure:inplanemovements3rows}(c)], this has no significant influence on the energy difference between the two different symmetry models. A histogram of all the bond lengths in the graphene superstructure demonstrates that the implementation of the lower $p3$-symmetry allows the bond lengths to be more preserved relative to bulk graphite.

We are aware of an independent low-energy electron-diffraction (LEED) study \cite{Moritz2010} on the same Ru single crystal using the same graphene preparation and characterization, where the authors claim a corrugation of the graphene layer of $1.5$~\AA, and a corrugation of the topmost ruthenium layer of $0.23$~\AA.  In that study the system is described by a $p3m1$-symmetry and the unit cell is cut down to one of the four inequivalent sub-unit cells in order to reduce computational time. The SXRD simulation of the coordinates extracted from the LEED-study led to a significantly higher R-factor of $34.0$~\%. The reason for the discrepancies, which are outside the error bars, are not yet resolved, although possible explanations are the already-mentioned restriction to $p3m1$-symmetry and a 12-on-11 superstructure -- a full dynamical scattering LEED calculation of the system with $p3$-symmetry is presently beyond computational capabilities. In addition, the fact that LEED only probes the topmost layers, while SXRD demonstrates that significant vertical displacements occur down to four atomic layers of the Ru-substrate, might also play an important role.

\section{Summary and conclusion}
In summary, we have determined the graphene/Ru(0001) structure in unsurpassed detail. This was only possible by adopting a parametric Fourier description of the superstructure using only a small number of physically reasonable parameters. Up to the mirror-symmetry breaking the final model agrees excellently with previous STM studies. We find a graphene and ruthenium corrugation peak-to-peak height of $(0.82\pm0.15)$~\AA\ and $(0.19\pm0.02)$~\AA, respectively. The ruthenium corrugation is out of phase with that of the graphene and decays exponentially down to a depth of several ruthenium layers. Importantly, we have also discovered the new and potentially highly significant property of areal chirality in the in-plane movements, which are most evident on the flanks of the hills of the corrugation. We propose that this symmetry-breaking phenomenon is induced by elastic energy minimization of the graphene layer. To test the validity of this, we calculated the elastic energy of the graphene superstructure to be $9.3$~eV, less than two thirds of that for the $p3m1$ case.

\section*{Acknowledgments}
Support of this work by the Schweizerischer Nationalfonds zur F\"orderurng der wissenschaftlichen Forschung and the staff of the Swiss Light Source is gratefully acknowledged. This work was performed at the Swiss Light Source, Paul Scherrer Institut.

\newpage

\appendix

\section*{Appendix}

\setcounter{section}{1}

\begin{figure}
  \includegraphics[width=1.0\columnwidth]{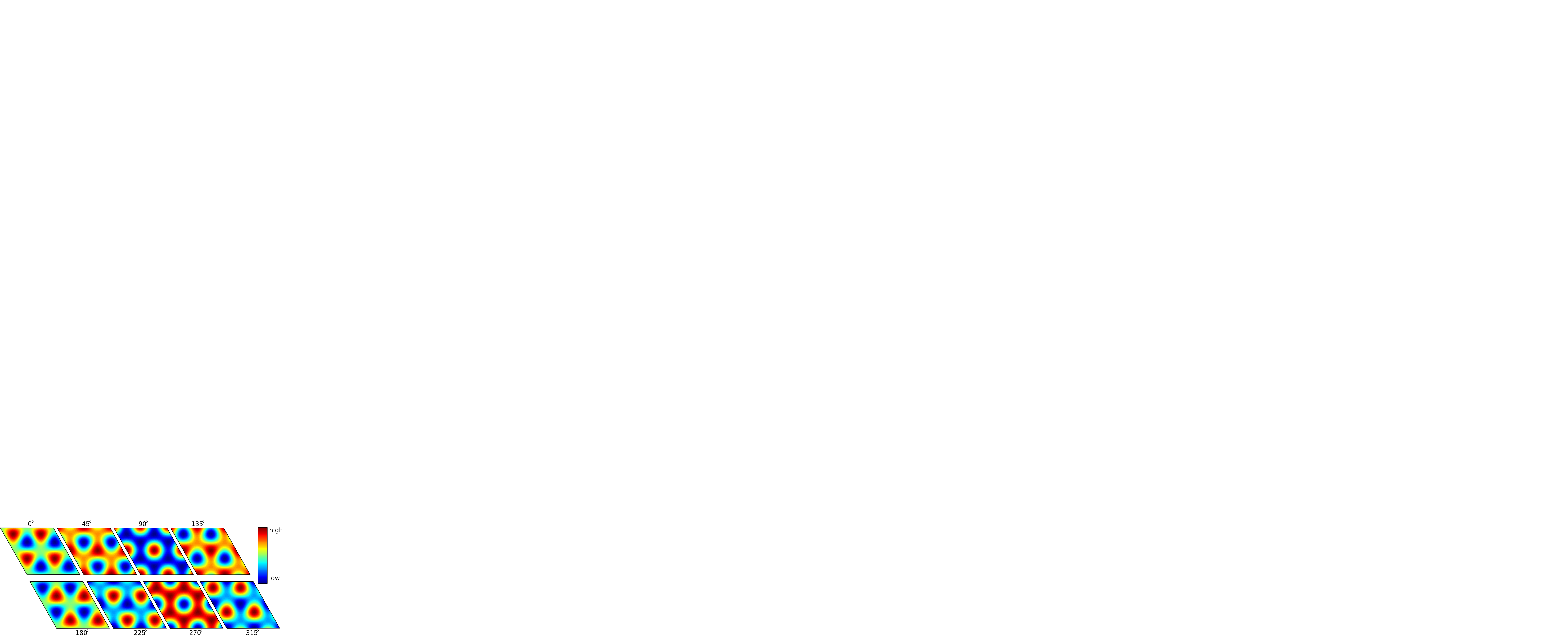}
  \caption{(color) Different corrugation shapes generated by different
  out-of-plane phase values. The blue regions are the strongly bound
  ``valleys'', the red highlighted regions show the weakly bound ``hills''.} 
  \label{figure:differentphases}
\end{figure}

In the following, the implementation of the symmetry constraints and the Fourier expansion to the graphene-on-ruthenium model will be briefly described. The displacement $\mathrm{d}\mathbf{r}$ of an atom sitting at point $\mathbf{r}$ is expressed by its 2-dimensional Fourier series
\begin{eqnarray}
  \mathrm{d}r^{i}         & = & \sum_{s,t} K_{s,t}^i\cdot \sin[2\pi(sx + ty) + \phi_{s,t}^{i}] \label{equation:fourier_series} \\
  K_{s,t}^{i}             & = & \sqrt{{A_{s,t}^{i}}^{2} + {B_{s,t}^{i}}^{2}}, \hspace{0.5cm} \phi_{s,t}^i  = \arctan(B^{i}/A^{i}) 
\end{eqnarray}
where $s$, $t \in \{0,2,4\}$ are the orders,  $A_{s,t}^{i}$, $B_{s,t}^{i}$ are the Fourier coefficients, $\phi_{s,t}^{i}$ are the phases of the corrugation and $i \in \{x,y,z\}$. Note that the phase of the out-of-plane displacements influences the valley and hill shapes and positions of the corrugation allowed by the $p3$ symmetry (Fig.~\ref{figure:differentphases}).

Since 
\begin{equation}
  \sin(f+\phi) = \sin(f)\cos(\phi)+\cos(f)\sin(\phi) \nonumber\\
\end{equation}
and equating $A^{i}$ and $B^{i}$ to
\begin{eqnarray}
  A^{i} & =  & K^{i}\cdot \cos(\phi^{i}) \nonumber\\
  B^{i} & =  & K^{i}\cdot \sin(\phi^{i}),
\end{eqnarray}
one can rewrite Eq.~\ref{equation:fourier_series} as
\begin{equation}
  \mathrm{d}r^{i} = \sum_{s,t} A_{s,t}^{i}\cdot \sin[2\pi(sx + ty)] + B_{s,t}^{i}\cdot \cos[2\pi(sx + ty)].
  \label{equation:2dfourierexpansion}
\end{equation}

\begin{figure}
  \includegraphics[width=1.0\columnwidth]{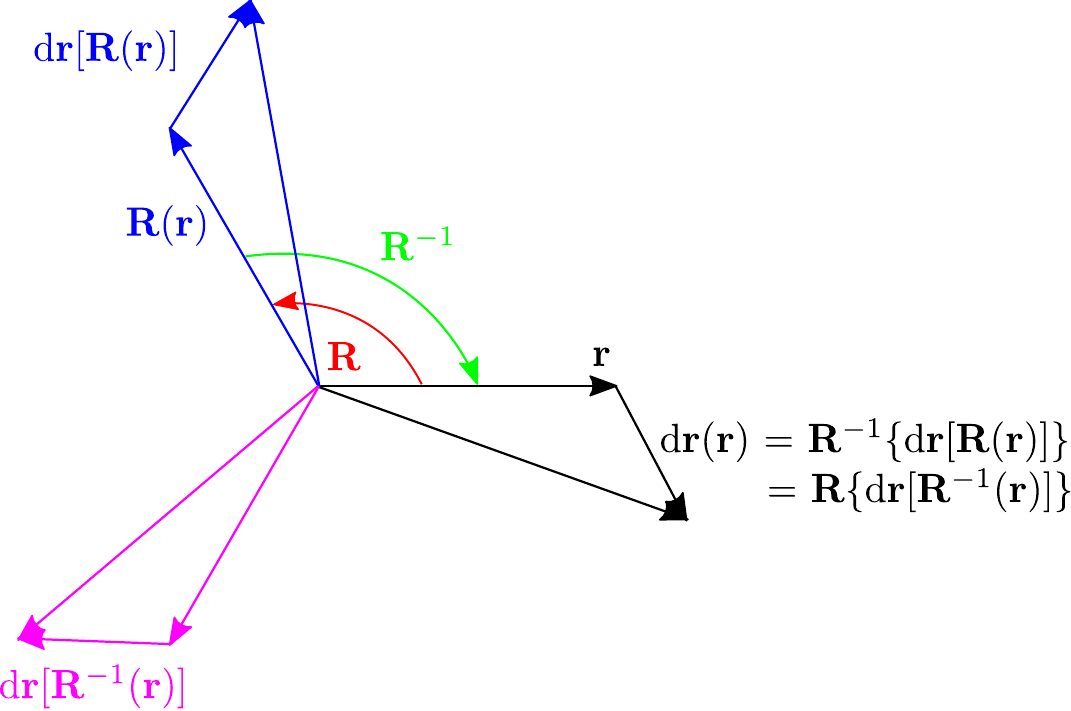}
  \caption{The $p3$-symmetry constraint operators. $\mathbf{R}$ is defined as a rotation by 120\degr \hspace{1mm} counterclockwise around the origin, while $\mathbf{R^{-1}}$ is the rotation clockwise by 120\degr \hspace{1mm} around the origin}
  \label{figure:p3symmetryconstraints}
\end{figure}

The rotation operators used for the description of the $p3$-symmetry are  $\mathbf{R}_{1}$  and $\mathbf{R}_{2}$, which in a hexagonal coordinate system describe a  $120$\degr~rotation counterclockwise and clockwise around the origin, respectively (Fig.\ref{figure:p3symmetryconstraints}), are given by
\begin{equation}
\mathbf{R}_{1}=\left(\begin{array}{cc} 0 & -1\\ 1&-1 \end{array}\right), \hspace{0.5cm} \mathbf{R}_{2}=\left(\begin{array}{cc} -1 & 1\\ -1 & 0 \end{array} \right). 
\label{equation:rotationmatrices}
\end{equation}

It can be easily shown that $\mathbf{R}_{1} = \mathbf{R}_{2}^{-1} \equiv \mathbf{R}$. $\mathrm{d}\mathbf{r}$ has to fulfill the $p3$-symmetry constraint, which results in 
\begin{equation}
\mathbf{R}^{-1}\{\mathrm{d}\mathbf{r}[\mathbf{R}(\mathbf{r})]\} = \mathbf{R}\{\mathrm{d}\mathbf{r}[\mathbf{R}^{-1}(\mathbf{r})]\} = \mathrm{d}\mathbf{r}(\mathbf{r}). 
\label{equation:fourierconstraints}
\end{equation}

The relations in Table~\ref{table:AandB} follow by inserting Eq.~\ref{equation:2dfourierexpansion} and Eq.~\ref{equation:rotationmatrices} in Eq.~\ref{equation:fourierconstraints}.

\begin{table}[h!]
\centering
  \begin{tabular}{lcrlcrl}
    \hline
    \hline	
    $A_{s,t}^{z}$ & = &       & $A_{t, -(s+t)}^{z}$  & = &     & $A_{-(s+t), s}^{z}$  \\   
    \hline
    $A_{s,t}^{x}$ & = &  $-$  & $A_{t, -(s+t)}^{x}  +  A_{t, -(s+t)}^{y}$   & = & $-$ & $A_{-(s+t), s}^{y}$   \\
    $A_{s,t}^{y}$ & = &       & $A_{-(s+t), s}^{x}  -  A_{-(s+t), s}^{y}$   & = & $-$ & $A_{t, -(s+t)}^{x}  $ \\
    \hline
    \hline
    $B_{s,t}^{z}$ & = &       & $B_{t, -(s+t)}^{z}$  & = &     & $B_{-(s+t), s}^{z}$  \\   
    \hline
    $B_{s,t}^{x}$ & = &  $-$  & $B_{t, -(s+t)}^{x}  +  B_{t, -(s+t)}^{y}$   & = & $-$ & $B_{-(s+t), s}^{y}$   \\
    $B_{s,t}^{y}$ & = &       & $B_{-(s+t), s}^{x}  -  B_{-(s+t), s}^{y}$   & = & $-$ & $B_{t, -(s+t)}^{x}  $ \\
    \hline
    \hline
  \end{tabular}
  \caption{The relations of the Fourier coefficients $A_{s,t}^{i}$ and $B_{s,t}^{i}$ ($i \in \{x,y,z\}$).}
  \label{table:AandB}
\end{table}

Regarding the considered Fourier components in the analysis, the zeroth order is the $23/23$~reflection, the first and third order components correspond to the $24/23$ and $26/23$ systematic absences, respectively, and the $25/23$ to the second order component. Hence the fourth order refers to the $27/23$-reflection, and along the $h$-direction (equivalent to the $k$-direction) one can limit ($s$,$t$) = $(2,0)$ and ($s$,$t$) = $(4,0)$. For the sake of simplicity, we describe here only the implementation of to the second order.

From Eq.~\ref{equation:2dfourierexpansion} and Table \ref{table:AandB}, one can derive the following expressions for the single components of  $\mathrm{d}\mathbf{r}$ which describe the displacement field. We do not include the orders $(s, t)$ for the sake of simplicity

\begin{eqnarray}
  \mathrm{d}r^{z}   & = &  A^{z}\cdot\sin(2\pi \cdot 2x) + A^{z}\cdot\sin[2\pi(-2y)] \nonumber\\
                    & + &  A^{z}\cdot\sin[2\pi (-2x+2y)] \nonumber\\
                    & + &  B^{z}\cdot\cos(2\pi \cdot 2x) + B^{z}\cdot\cos[2\pi(-2y)]\nonumber\\
                    & + &  B^{z}\cdot\cos[2\pi (-2x+2y)]\\
  \mathrm{d}r^{x}   & = &  A^{x}\cdot\sin(2\pi \cdot 2x) - A^{y}\cdot\sin[2\pi(-2x+2y)]\nonumber\\
                    & + & (A^{y} - A^{x})\cdot\sin[2\pi \cdot(-2y)] \nonumber\\
                    & + &  B^{x}\cdot\cos(2\pi \cdot 2x) - B^{y}\cdot\cos[2\pi(-2x+2y)]\nonumber\\
                    & + & (B^{y} - B^{x})\cdot\cos[2\pi \cdot(-2y)]\\
  \mathrm{d}r^{y}   & = &  A^{y}\cdot\sin(2\pi \cdot 2x) - A^{x} \cdot\sin[2\pi \cdot(-2\pi\cdot2y)]\nonumber\\
                    & + & (A^{x} -A^{y})\cdot\sin[2\pi(-2x+2y)]\nonumber\\
                    & + &  B^{y}\cdot\cos(2\pi \cdot 2x) -  B^{x} \cdot\cos[2\pi \cdot(-2\pi\cdot2y)]\nonumber\\
                    & + & (B^{x} -B^{y})\cdot\cos[2\pi(-2x+2y)]. 
\end{eqnarray} 

\begin{figure}
  \includegraphics[width=1.0\columnwidth]{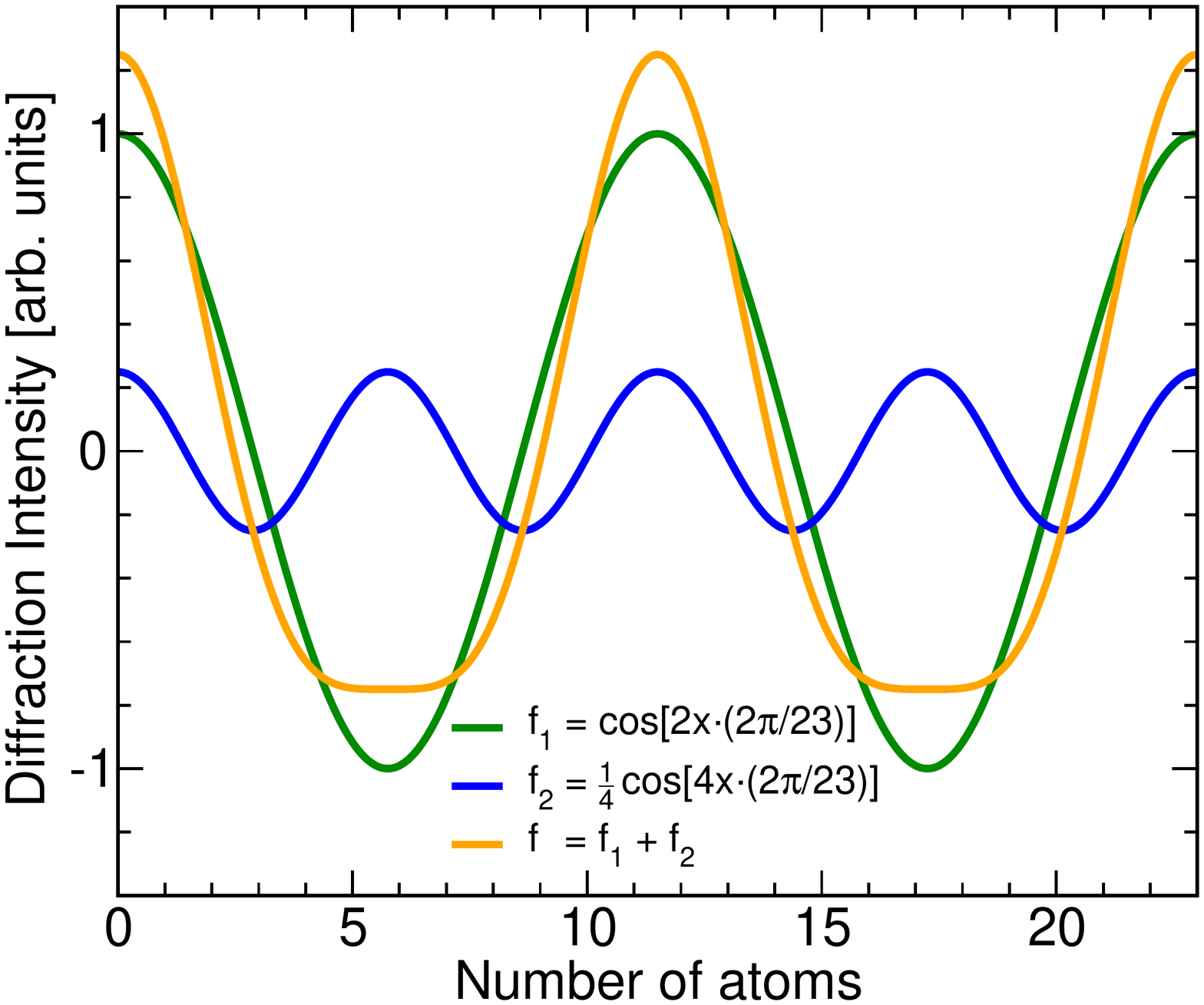}
  \caption{Effect of the implementation of the fourth harmonic: $f_{1}$ represents the second harmonic, $f_{2}$ the fourth. Their sum for an amplitude of the fourth harmonic up to $0.25$ of that of the second harmonic makes the low regions flatter.}
  \label{figure:fourthharmonic}
\end{figure}
So, we obtain six fitting parameters for the displacement field $\mathrm{d}\mathbf{r}$, namely $A^{x}$, $A^{y}$, $A^{z}$, $B^{x}$, $B^{y}$, $B^{z}$. Since we lock the phase for the fourth order to be the same as that for the second order, there will be nine fitting parameters. The effect the fourth order harmonic with a locked phase has on the structure is shown in Fig. \ref{figure:fourthharmonic}. For an amplitude up to $0.25$ of that of the second harmonic, the low regions in the structure will be flattened out.

\newpage

\section*{References}
\bibliographystyle{unsrt}
\bibliography{citation_list}

\end{document}